\documentclass[
reprint,
showpacs,preprintnumbers,
showkeys,
amsmath,amssymb,
aps,
prapplied,
longbibliography,
floatfix,superscriptaddress
]{revtex4-2}

\usepackage[english]{babel}
\usepackage[utf8]{inputenc}
\usepackage{graphicx}
\usepackage[colorlinks=true,linkcolor=blue,citecolor=blue,urlcolor=blue]{hyperref}
\usepackage{xcolor}
\usepackage{amsmath}
\usepackage{amsfonts}
\usepackage{amssymb}
\usepackage{dsfont}
\usepackage{siunitx}
\usepackage{xcolor,import}
\usepackage{transparent}
\usepackage{sidecap} 
\usepackage{newtxtext,newtxmath}
\usepackage[T1]{fontenc}
\usepackage{subcaption}
\usepackage{comment}
\usepackage{natbib}
\usepackage[title,titletoc]{appendix}

\graphicspath{Figures/}

\DeclareSIUnit{\gauss}{G}   
\DeclareSIUnit{\bar}{bar}     

\definecolor{refColor}{HTML}{EA00F2}
\definecolor{figColor}{HTML}{008DF2}
\definecolor{urlColor}{HTML}{00AEF2}

\hypersetup{
    unicode=false,                 
    pdftoolbar=true,               
    pdfmenubar=true,               
    pdffitwindow=false,            
    pdfstartview={FitH},           
    pdftitle={},                   
    pdfsubject={Quantum physics},  
    pdfkeywords={},                
    pdfnewwindow=true,             
    colorlinks=true,               
    linkcolor=figColor,            
    citecolor=refColor,            
    filecolor=magenta,             
    urlcolor=urlColor              
}

\begin{document}

\title{Two-color Ytterbium MOT in a compact dual-chamber setup}

\author{Xin Wang}
\thanks{These authors contributed equally to this work.}
\author{Thilina Muthu-Arachchige}
\thanks{These authors contributed equally to this work.}
\author{Tangi Legrand}
\author{Ludwig Müller}
\author{Wolfgang Alt}
\author{Sebastian Hofferberth}
\email{hofferberth@iap.uni-bonn.de}
\author{Eduardo Uruñuela}
\affiliation{Institute of Applied Physics, University of Bonn, Bonn, Germany}

\date{\today}


\begin{abstract}
We present an experimental scheme for producing ultracold Ytterbium atoms in a compact dual-chamber setup. A dispenser-loaded two-dimensional (2D) magneto-optical trap (MOT) using permanent magnets and operating on the broad $^1S_0\to {}^1P_1$ singlet transition delivers over $10^7$ atoms per second through a differential pumping stage into a three-dimensional (3D) MOT. The two-color 3D MOT uses the broad singlet transition to accumulate $\sim\!2\times 10^7$ atoms of $^{174}\text{Yb}$ within \qty{2.5}{\second} and subsequently the narrow $^1S_0\to {}^3P_1$ intercombination line to cool the atomic cloud to below \qty{10}{\micro\kelvin}. We report optimized parameters for each stage of the atom collection sequence, achieving high transfer efficiency. We find that shelving into the triplet state during the broad-transition MOT almost doubles the number of trapped atoms.

\end{abstract}

\maketitle

\section{\label{sec 1:level 1: Introduction}Introduction}
In the field of cold atoms, there has been a rapid growth of interest in two-valence-electron species~\cite{first_AEML_atoms_trapping_Shimizu, Sr_first_trapping_Katori, Yb_first_MOT_Kuwamoto, Sr_BEC_Schreck, AEML_atoms_quantum_computing_Zoller, Rydberg_with_AEML_atoms_TPohl, AEML_atoms_quantum_computing_Fross-Seig, AEML_atoms_Rydberg_Endres}. These divalent atoms possess a richer energy level structure compared to single-valence-electron species, offering broad, narrow and ultra-narrow linewidth optical transitions, isotopes with spinless ground-states or purely nuclear hyperfine levels. Therefore, such atoms are used in numerous applications including optical atomic clocks~\cite{Yb_atomi_clock_Ludlow, Sr_lattice_clock_Ye, Sr_optical_clock_Endres, Sr_Lattice_clock_Endres}, atomic gravimeters~\cite{Sr_Gravimeter_Tino}, interferometers~\cite{Yb_Interferometry_Gupta}, superradiant lasers~\cite{Sr_Superradiance_laser_JKThompson}, quantum simulations~\cite{Sr_Lattices_Dalibard, Sr_Optical_tweezers_Endres}, quantum computing~\cite{AEML_atoms_quantum_computing_Fross-Seig, AEML_atoms_quantum_computing_Zoller, Sr_Quantum_computing_Buchler}, and molecular physics~\cite{Sr_molecular_physics_Hutson, Sr_molecular_physics_Zelevinsky}. In the context of Rydberg physics experiments, the presence of two valence electrons may offer several advantages~\cite{Rydberg_two_electron_dynamic_Jones, AEML_atoms_Rydberg_Endres}. 
For instance, once an atom is excited into a Rydberg state, the presence of the second valence electron may facilitate optical imaging of the Rydberg atoms~\cite{Yb_Rydberg_Tweezer_array_Takahashi}, and simultaneous trapping of ground state and Rydberg state atoms~\cite{Rydberg_with_AEML_atoms_TPohl}. For Ytterbium, two-photon collective Rydberg excitations with closely-spaced wavelengths of probe (\qty{\sim399}{\nano\metre}) and control (\qty{\sim395}{\nano\metre}) promise reduced motional dephasing rates~\cite{li_entanglement_2013}.

So far, alkaline-earth and alkaline-earth-like atom experiments have utilized separate sources to feed atoms into a magneto-optical trap (MOT), often realized as an effusive oven plus a Zeeman slower. To efficiently load large numbers of atoms while maintaining ultra-high vacuum in a separate science chamber, several approaches such as mechanical shutters~\cite{cheiney_zeeman_2011} and optical deflectors~\cite{Sr_MOT_deflector_and_Zeeman_Willkowski, AOSense_2024} have been integrated into Zeeman slower-based setups. A compact permanent-magnet-based Zeeman slower has been demonstrated~\cite{Sr_Zeeman_slower_Permanent_Magnets_Gill} and recently combined with an additional deflector stage and a 2D MOT \cite{Yb_MOT_PermanentMagnet_Zeeman_Schlippert}. Alternative approaches such as effusive ovens with capillaries~\cite{Yb_oven_with_capilaries_Kaiser}, effusive oven-based 2D MOTs~\cite{Sr_2DMOT_with_Oven_Weidemuller, Dy_2DMOT_with_Oven_Chomaz, barbiero_sideband-enhanced_2020}, and dispenser-loaded 2D MOTs~\cite{Yb_first_dispenser_2DMOT_Dorscher, Sr_dispenser_loaded_2DMOT_Will} have also been demonstrated recently. 

Various concepts have also been developed for cooling and trapping two-valence-electron atoms in 3D MOTs taking advantage of their narrow-linewidth transitions~\cite{Sr_first_trapping_Katori, Yb_first_MOT_Kuwamoto}. In the simplest approach, atoms can directly be loaded into a narrow-linewidth MOT, using frequency broadening to increase the capture velocity~\cite{Yb_first_MOT_Kuwamoto, Yb_first_dispenser_2DMOT_Dorscher}.
Spatially separated pre-slowing of atoms via the broad transition has also been employed to reduce the requirements on capture velocity of the narrow-line MOT~\cite{Yb_slowing_beams_Gupta, Er_slowing_beams_Gyo-boong, Dy_slowing_beams_Ketterle}. 
As an alternative, two-color MOTs have been implemented in different variants. The broad- and narrow-linewidth MOTs can be operated simultaneously in an overlapping form~\cite{Yb_Two_color_MOT_Vuletic}, spatially separated by arranging them in a core-shell configuration~\cite{Yb_core-shell_MOT_Mun} or temporally separated~\cite{Sr_first_trapping_Katori, Sr_BEC_Schreck} by first loading the atoms into a broadband MOT and subsequently handing over to a narrow-linewidth MOT.

In this paper, we present a compact double-MOT setup for preparing cold ensembles of $^{174}\text{Yb}$ atoms, which we use for Rydberg-mediated nonlinear quantum optics experiments~\cite{firstenberg_nonlinear_2016, murray_chapter_2016, pritchard_cooperative_2010, dudin_strongly_2012, peyronel_quantum_2012}. The double-MOT setup consists of a dispenser-loaded 2D MOT loading a two-color 3D MOT. The 3D MOT utilizes three sequential stages in time: a broad-linewidth MOT stage for accumulating atoms, an efficient transfer stage, and a final narrow-linewidth MOT to achieve high atom numbers at low temperatures. Optical shelving during the broad-line MOT increases the final number of trapped atoms by nearly a factor of two. After the thorough optimization of all stages, our setup produces a sample of $\sim\! 1.8\times10^7$ atoms of $^{174}\text{Yb}$ at a temperature below \qty{10}{\micro\kelvin} in \qty{2.5}{} seconds.

\section{\label{sec2:level 1: experimental setup}Experimental Setup and Sequence}
\begin{figure*}
    \includegraphics[scale = 1]{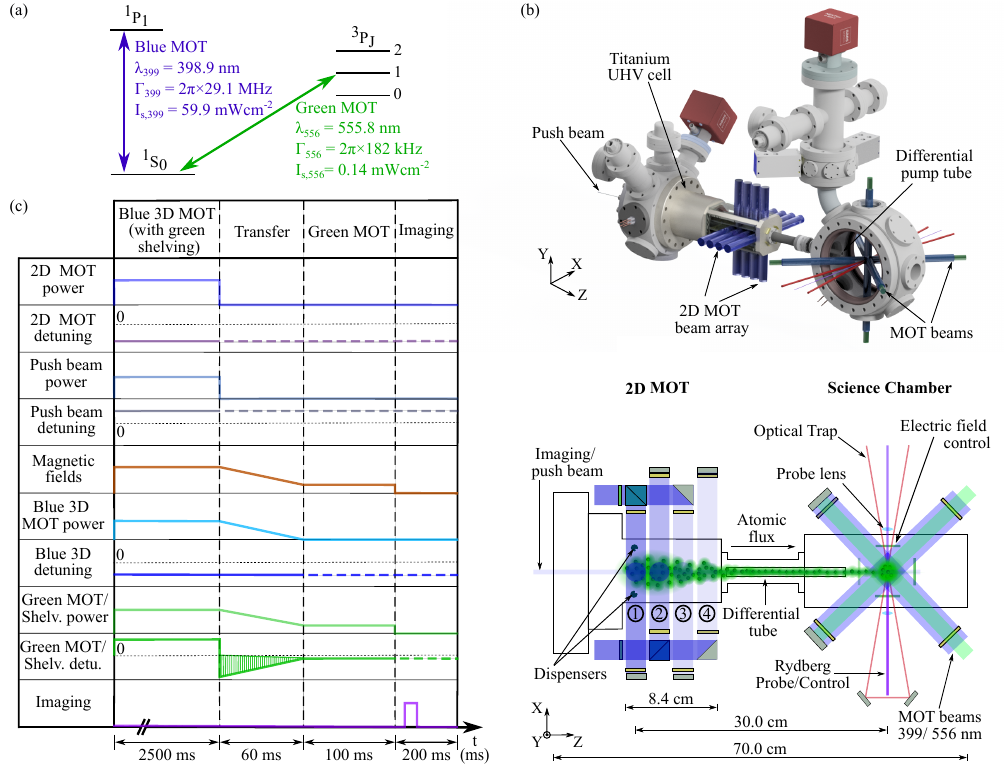}
    \caption{\label{fig:2D-3D view of the setup} 
        Experimental setup and sequence. (a) Laser cooling transitions of $^{174}\text{Yb}$. (b) Top: CAD model of the dual-chamber vacuum apparatus including the laser beams used for the 2D MOT, two-color 3D MOT, dipole trap and Rydberg excitation. Bottom: 2D schematic of the setup as viewed from above. (c) Two-color MOT loading sequence with broad-linewidth blue MOT, transfer, and narrow-linewidth green MOT stages. Absorption imaging with time-of-flight is used to measure the atom number and temperature of the final green MOT.
    } 
\end{figure*}
The two relevant transitions for laser-cooling $^{174}\text{Yb}$ are shown in Figure~\ref{fig:2D-3D view of the setup}(a), namely the broad \emph{blue} singlet transition $^1S_0\to {}^1P_1$ at a wavelength of \qty{399}{\nano\metre} and the narrow \emph{green}  intercombination line $^1S_0 \to {}^3P_1$ at \qty{556}{\nano\metre}.

Our vacuum system consists of a 2D MOT chamber and a science chamber separated by a differential pumping tube. Figure~\ref{fig:2D-3D view of the setup}(b) shows a three-dimensional computer-aided design (CAD) model of the vacuum system and a two-dimensional schematic of the top view of the setup. 
The 2D MOT chamber comprises an 8~inch commercial octagon and an in-house designed ultra-high vacuum (UHV) cell enclosing the Yb dispenser system. The UHV cell is made of anti-reflection coated BK7 glass windows glued onto a Titanium frame with UHV-compatible glue (EPO-TEK H77S), and supports pressures of \qty{e-10}{\milli\bar}. 
The science chamber consists of an 8~inch commercial octagon, three 4~inch windows, and two large 8~inch UV-fused silica viewports for good optical access that are anti-reflection coated for wavelengths \qtylist{395;399;556;1064}{\nano\meter}. In order to facilitate future Rydberg quantum optics experiments, the science chamber hosts a set of electrodes for controlling electric fields and a multi-channel-plate ion detector. Additionally, a crossed-beam optical dipole trap and Rydberg excitation beams will be added to the setup, as illustrated in Fig.~\ref{fig:2D-3D view of the setup}(b). The differential pumping tube connecting the 2D MOT chamber and the science chamber --with \qty{16.5}{\centi\meter} length and \qty{3}{\milli\meter} diameter-- isolates the science region from background gas emitted by the high-temperature dispensers, allowing a differential pressure of two orders of magnitude. The outer dimensions of the entire apparatus are \qtyproduct{70 x 55 x 69}{\centi\meter}.

The 2D MOT uses an array of four intersecting Gaussian beams driving the blue transition. A two-dimensional quadrupole field is provided by permanent magnets. A weak push beam, aligned through the 2D MOT and the differential pumping tube, enhances the atomic flux into the science chamber. The full details are presented in Sec.~\ref{sec4:level 1: Yb atom source}.

The 3D MOT uses both the blue transition as well as the green transition. Such two-color MOT operation is separated in time into three distinct stages: blue MOT, blue-green transfer, green MOT, as shown in Fig.~\ref{fig:2D-3D view of the setup}(c). The experiment starts with the blue MOT loading, where the 2D MOT, push beam, and blue MOT operate together for \qty{2500}{\milli\second}. To increase the atom number in the blue MOT, we shelve a fraction of the atoms into the long-lived ${}^3P_1$ triplet state by shining green laser light with positively-detuned frequency onto the atoms, see Sec.~\ref{sec5:level 1: Optical shielding of Blue 3D MOT}. Next, the \qty{60}{\milli\second} long blue-green transfer stage requires ramping down the magnetic field gradient and optical power of the blue and green beams to their optimum values, while the green MOT light is frequency-broadened, as shown in Fig.~\ref{fig:2D-3D view of the setup}(c), to improve its capture volume and velocity, see Sec.~\ref{sec:transfer}. After the transfer, atoms are held for \qty{100}{\milli\second} by a pure green MOT without frequency broadening to reach equilibrium at low temperature, see Sec.~\ref{sec:GreenMOT}. 

Finally, in the last \qty{200}{\milli\second}, absorption imaging with time-of-flight expansion is used to determine the number of trapped atoms $N_\text{A}$ and their temperature $T_A$ in the 3D green MOT. Our setup allows imaging on the blue transition along the $y$ and $z$ axes as illustrated in the lower panel of Fig.~\ref{fig:2D-3D view of the setup}(b). 
In the 3D blue MOT, the atom cloud is too large and dilute for absorption imaging. Therefore, we determine the loading dynamics of the blue MOT from the intensity of the blue fluorescence light $F(t)$ over time, recorded by an amplified photodiode. We use both absorption and fluorescence signals in the following sections for the optimization of the different stages of our sequence.

\section{\label{sec4:level 1: Yb atom source}Blue 2d MOT} 
\begin{figure}
    \includegraphics[scale = 1]{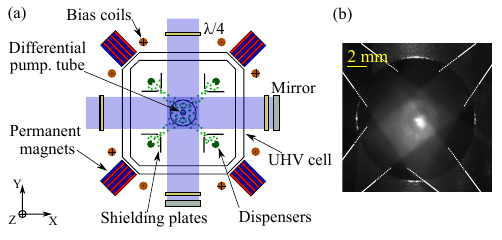}
    \caption{\label{fig:2DMOT Cross Section} 
        Our 2D MOT setup. (a) Schematic of the 2D MOT cross section showing the permanent magnets, four Yb dispensers and retro-reflecting, crossed Gaussian beam pairs, centered on the differential pumping tube. (b) Fluorescence of the chain of four 2D-MOTs imaged through the rear viewport. The dispenser emission cones can be seen as weak fluorescence, outlined by the white dashed lines.  
    }
\end{figure}
\begin{figure}
    \includegraphics[scale = 1]{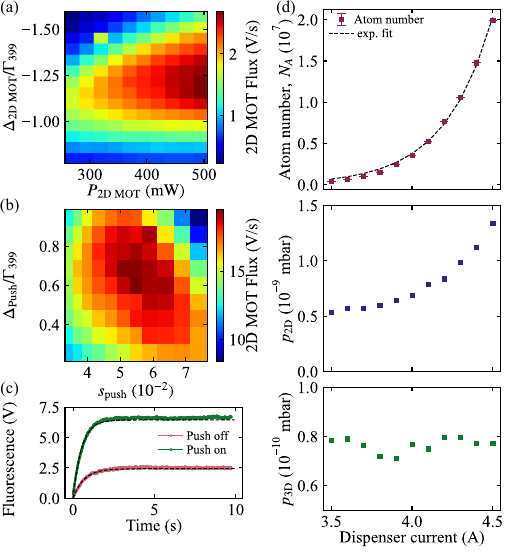} 
    \caption{\label{fig:2D MOT performance}
        Optimization of the 2D MOT. (a) Estimated 2D MOT flux as function of the total power $P_\text{2D MOT}$ and detuning $\Delta_\text{2D MOT}$ of the 2D MOT. Push beam parameters are $\Delta_\text{push} = 0.6\times\Gamma_\text{blue}$ and $s_\text{push} = 5.5\times10^{-2}$. (b) Estimated 2D MOT flux as a function of saturation parameter $s_\text{push}$ and detuning $\Delta_\text{push}$ of the push beam.  (c) Loading curve of the blue 3D MOT with and without the push beam, showing an increase of steady-state fluorescence by a factor \qty{\sim3}{}. The 2D MOT parameters are $P_\text{2D MOT} = \qty{360}{\milli\watt}$  and $\Delta_\text{2D MOT} = -1.2\times\Gamma_\text{blue}$ for (b) and (c). (d) Atom number (top) in the green MOT, pressure in the 2D MOT chamber (middle) and the science chamber (lower) as a function of dispenser current.}
\end{figure}
Our setup uses Yb dispensers (Alfavacuo S-type AS-3-Yb-0500) to feed the 2D MOT which in turn creates an atomic beam propagating into the science chamber. A cross-section of our 2D MOT setup is shown in Figure~\ref{fig:2DMOT Cross Section}(a).  Four dispensers are mounted symmetrically around the center axis of the 2D MOT cell. To minimize the Yb deposition on the optical viewports of the UHV cell, we mount L-shaped shields around each dispenser.
The 2D MOT trapping beams are all derived from one same blue laser beam with a $1/e^2$ radius of \qty{9}{\milli\meter}, which is divided into eight independent beams aligned to form four crossed retro-reflected beam pairs. This 2D MOT beam array creates four separate loading zones, i.e. a chain of four 2D-MOTs, as shown in Figure~\ref{fig:2D-3D view of the setup}(b). 
To produce a 2D magnetic quadrupole field in the $xy$-plane, we use four stacks of neodymium permanent magnets (Eclipse Magnetics N750-RB). The stacks of 28 magnets each are mounted around the 2D MOT chamber as illustrated in Figure~\ref{fig:2DMOT Cross Section}. By radially sliding the magnet stacks in their mounts we can produce gradients from \qtyrange{20}{50}{\gauss\per\centi\meter}. The magnet arrangement has been designed to maintain a transverse field gradient constant to within $2.5\%$ over the full length of the four loading zones along the $z$-direction. The transverse center of the 2D quadrupole field  and the center of the four crossed beam array are aligned with the differential pumping tube. To facilitate the fine alignment of the 2D-MOTs axis with the differential pumping tube, two pairs of compensation coils generate bias fields of \qty{4.2}{\gauss/\ampere} in the $xy$-plane. 

Figure.~\ref{fig:2DMOT Cross Section}(b) shows the fluorescence of Yb atoms during the operation of the 2D MOT, as observed through the rear viewport. The emission cones of the dispensers are highlighted by the white dashed lines. The bright spot in the center corresponds to the chain of four 2D-MOTs aligned to the center of the differential pumping tube. 

To maximize the atomic flux into the science chamber, we optimize the magnetic field gradient and the optical power distribution between the four 2D-MOTs. We find the optimal gradient to be \qty{34}{\gauss\per\centi\meter}, and the optimum optical power distribution to be imbalanced with $\sim60\%$ in 2D-MOT-1 (furthest from the science chamber), $\sim20\%$ in both the following 2D-MOTs, and $\sim1\%$ in 2D-MOT-4 (closest to the science chamber). As 2D-MOT-1 is closest to the emission slit of the dispensers, it captures the majority of atoms. Then 2D-MOT-2 and 2D-MOT-3 load additional atoms and guide the transversely cold atoms from 2D-MOT-1 towards the science chamber. Lastly, we observe that while 2D-MOT-4 needs very low power relative to the other three, it is critical for steering the atomic beam through the differential pumping tube.

With 2D MOT magnetic fields and power distribution optimized, we investigate the dependence of the 2D MOT flux on the total optical power $P_\text{2D-MOT}$ and the detuning $\Delta_\text{2D-MOT} = \omega_\text{2D-MOT} - \omega_\text{399}$ of the blue 2D MOT light, as shown in Fig.~\ref{fig:2D MOT performance}(a). To determine the atomic flux from the measured blue MOT fluorescence $F(t)$, we fit loading curves of the form $F(t)=F_0 (1-e^{-\alpha t})$ to the recorded data, and extract the loading rate $\mathrm{\alpha}$ and the steady state amplitude $F_0$. The initial slope $\alpha F_0$ of the loading curve is proportional to the captured atomic flux from the 2D MOT and is used to optimize the 2D MOT performance.
The atomic flux peaks at a detuning of $\Delta_\text{2D-MOT} \!\approx\! -1.2\times\Gamma_{\text{blue}}$ and at the highest optical power of $P_\text{2D-MOT}\!\approx\!\qty{490}{\milli\watt}$, indicating that the 2D MOT performance is limited by the available optical power. Our experimental data is in good agreement with numerical simulations of the 2D MOT, which we present in Appendix~\ref{sec:MOT-simulation}. 

The addition of the blue push beam with $1/e^2$ radius of \qty{0.9}{\milli\meter} significantly increases the atomic flux into the science chamber. In comparison, we find that a green push beam as used in Ref.~\cite{Dy_2DMOT_with_Oven_Chomaz} provides no significant improvement in our configuration. 
In Fig.~\ref{fig:2D MOT performance}(b), we show the dependence of the flux on push beam detuning $\Delta_\text{push}$ and saturation parameter $s_\text{push}$, controlled independently from the parameters of the MOT light. We observe maximal flux for positive detuning, peaking at $\Delta_\text{push} = 0.6\times\Gamma_\text{blue}$ and $s_\text{push} = 0.05$. We conclude from this that the push beam accelerates slowly moving atoms to the minimum velocity of \qty{\sim10}{\meter\per\second} required to pass through the differential pumping tube. To quantify the flux increase, we compare in Fig.~\ref{fig:2D MOT performance}(c) loading curves with and without push beam.  We find an enhancement by a factor of $\sim\!3$ with the push beam at optimal parameters.

Furthermore, we investigate the final number of trapped atoms $N_\text{A}$ in the 3D MOT for different dispenser currents, as shown in Fig.~\ref{fig:2D MOT performance}(d). We find an approximately exponential rise of $N_\text{A}$ with increasing dispenser current. This is  accompanied by a corresponding rise of pressure in the 2D MOT chamber, while the science chamber pressure remains stable at \qty{\sim8e-11}{\milli\bar}. This indicates that we could apply higher dispenser currents to reach larger number of atoms without affecting the pressure in the science chamber. To extend the lifetime of the dispensers, we choose to operate them at \qty{4.2}{\ampere}, yielding a power consumption of \qty{9.6}{\watt}.

\section{\label{sec5:level 1: Optical shielding of Blue 3D MOT} Blue 3D MOT with shelving enhancement}
\begin{figure}
    \centering
    \includegraphics[scale = 1]{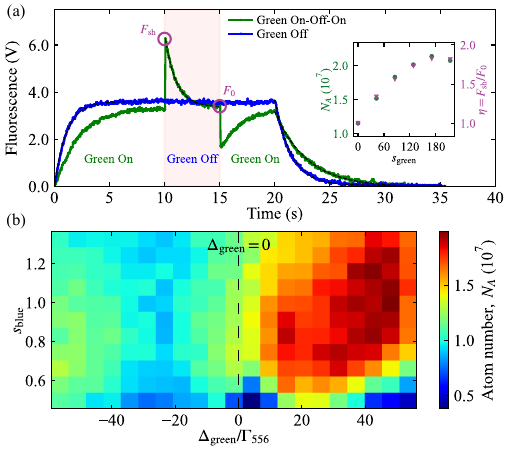}
    \caption{\label{fig:Optical shelving in Blue MOT}
        Optical shelving with green light during the blue MOT loading. (a) Comparison of fluorescence loading curves of the blue MOT with the green beams off and intermittently on-off-on, with $s_\text{blue} = 1.3$. The inset shows the atom number $N_\text{A}$ in the final green MOT and the enhancement factor of the blue fluorescence $\eta = F_\text{sh}/F_{0}$ due to shelving. (b) Atom number $N_\text{A}$ as a function of detuning of the green beams $\Delta_\text{green}$ from the bare atomic resonance $\Delta_\text{green}=0$ during the shelving, showing the influence of the blue MOT saturation parameter $s_\text{blue}$. In (a) and (b) $B'_y = \qty{12}{\gauss\per\centi\meter}$ and $\Delta_\text{blue} = -1.2\times\Gamma_{399}$.
    }
\end{figure}

In the science chamber, three pairs of counter-propagating blue laser beams with a $1/e^2$ radius of \qty{7.5}{\milli\meter} form the 3D blue MOT (see Fig.~\ref{fig:2D-3D view of the setup}(b)). The beams in the $xz$-plane have twice the power of those along the $y$ axis to compensate for the half as strong magnetic field gradient in the $xz$-plane. A pair of coils in  anti-Helmholtz configuration generates a quadrupole field with a maximum gradient of $B'_y=\qty{43}{\gauss\per\centi\meter}$. Additionally, three pairs of Helmholtz coils mounted along the $x, y$, and $z$ axes are used to compensate for any residual magnetic fields. 
We find that the number of atoms loaded into the blue MOT significantly increases when simultaneously strong green light is applied with positive detuning from the intercombination line. For this we use the green MOT beams which are superimposed on the blue MOT beams, with a $1/e^2$ radius of \qty{6}{\milli\meter} and balanced power.
We interpret this effect as optical shelving into the long-lived ${}^3P_1$ state, as has been recently observed in a Strontium MOT~\cite{Sr_MOT_ICFO}.
Optical shelving reduces atom loss due to light-induced collisions by transferring part of the atomic population into a long-lived state which is decoupled from the strong cooling transition. This effect is analogous to the idea behind dark-spot MOTs for alkali atoms~\cite{ketterle_high_1993}.

\begin{figure}
    \centering
    \includegraphics[scale = 1]{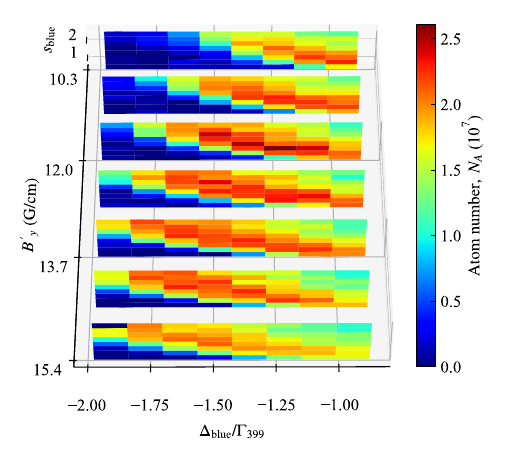}
    \caption{\label{fig:3D MOT characterization}
        Optimization of the blue MOT with shelving. Atom number $N_\text{A}$ in the final green MOT, as a function of the detuning $\Delta_\text{blue}$, the magnetic field gradient $B'_y$ and the saturation parameter $s_\text{blue}$ in the blue MOT.
    }
\end{figure}

To quantify the influence of green shelving on the blue MOT performance, we record  blue fluorescence loading curves with the green beams off and intermittently on-off-on, as shown in Fig.~\ref{fig:Optical shelving in Blue MOT}(a).  
After initial loading to saturation with shelving, when switching off the green light, the shelved atoms decay back to the ground state within the lifetime of the ${}^3P_1$ state and enter the cooling cycle. The fluorescence sharply increases to $F_\text{sh}$ beyond the steady-state fluorescence of the MOT without shelving, which shows that significantly more atoms have been accumulated but that their fluorescence was suppressed by the shelving. This higher number of unshelved atoms are then subject to increased light-induced losses and the fluorescence relaxes back to the equilibrium level $F_0$ of the unshelved MOT. 
When the green beams are turned on again, the suppression of fluorescence by almost a factor of 2 is clearly visible, which is expected in the strong saturation limit of the green transition.
At the \qty{20}{\second} mark we turn off the 2D MOT and observe the decay of the atom number. The significantly slower decay of the shelved MOT confirms that shelving reduces atom loss.

From the extracted fluorescence values of $F_\text{sh}$ and $F_0$, we estimate the shelving enhancement factor $\eta = F_\text{sh}/F_{0}$. In the inset of Figure~\ref{fig:Optical shelving in Blue MOT}(a) we show $\eta$, together with the final atom number $N_\text{A}$ in the green MOT, as a function of the saturation parameter $s_\text{green}$ which is proportional to the total power of the green beams. 
The final atom number $N_\text{A}$ and $\eta$ have the same dependency on the green saturation parameter, and reach an upper limit at $\eta\approx1.8$ and $N_\text{A} \approx2.3\times10^7$.

Additionally, we investigate the frequency dependence of the shelving effect and its interplay with the blue MOT power. Figure.~\ref{fig:Optical shelving in Blue MOT}(b) depicts the atom number $N_\text{A}$ as a function of the blue MOT saturation parameter $s_\text{blue}$ and green light detuning $\Delta_\text{green} = \omega_\text{green} - \omega_\text{556}$. The atom number peaks for positive detunings of the green light with respect to the bare atomic resonance $\Delta_\text{green}=0$ denoted by the dashed line. This optimum shifts to larger positive detunings with increasing blue MOT intensity.
We interpret the line shift as the AC-Stark shift of the ground state $^1S_0$ induced by the near-resonant blue light. The line broadening is attributed to the inhomogeneous intensity distribution of the blue light, the spatial distribution of the atomic cloud and the Zeeman shift of the upper $^3P_1$ substates in the MOT quadrupole field, as observed for Strontium in \cite{Sr_MOT_ICFO}. Estimates of the AC Stark shift based on the dressed state model for the blue singlet transition, and the Zeeman shift of the green triplet transition averaged over the atomic cloud, are in qualitative agreement with the experimental observations.
A more sophisticated model would need to take the atomic distribution and local polarization of the green light as well as local magnetic field direction into account.
Our observations indicate that the shelving works best on resonance with the \emph{light-shifted} narrow green transition.

We then optimize the atom number $N_\text{A}$ as a function of the blue MOT parameters with green shelving, namely saturation parameter $s_\text{blue}$, detuning $\Delta_\text{blue} = \omega_\text{blue} - \omega_\text{399}$, and magnetic field gradient $B'_y$, as shown in Figure~\ref{fig:3D MOT characterization}.
 We observe a shift of the best detuning $\Delta_\text{blue}$ towards negative values when the magnetic field gradient is increased, as expected for a MOT. We determine the overall best combination of parameters to be $s_\text{blue} \approx 0.35$, $\Delta_\text{blue} = -1.3\times\Gamma_{399}$, and $B'_y = \qty{12.0}{\gauss\per\centi\meter}$, yielding $N_\text{A} \sim 2.4\times 10^7$ atoms in the final green MOT. The absolute maximum lies within the three-dimensional parameter space reachable in the experiment, indicating that the atom number is not limited by any of the parameters.

\section{\label{sec:transfer}Blue-Green MOT transfer}

\begin{figure}
    \centering
    \includegraphics[scale = 1]{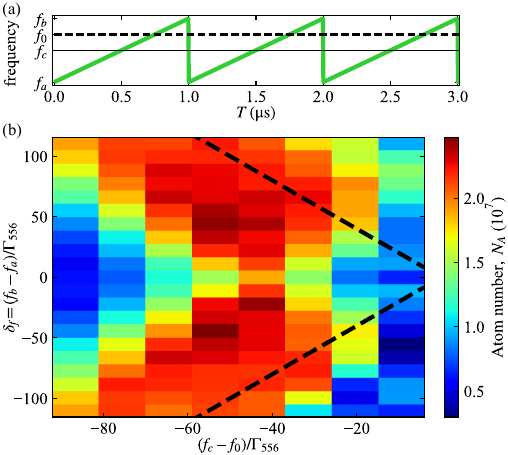}
    \caption{\label{fig: Green MOT transfer}
        Optimization of the transfer stage. (a) Sawtooth frequency modulation used to broaden the frequency of the green MOT light, defined by the center frequency $f_c$, start frequency $f_a$ and stop frequency $f_b$. The dashed line at $f_0$ denotes the atomic resonance frequency. (b) Atom number $N_\text{A}$ in the green MOT as a function of the detuning of the center frequency $f_c - f_0$ and the range of the sweep $\delta_f =f_b - f_a$. The black dashed lines show the condition $f_c - f_0 = -|\delta_f /2|$.
    }
\end{figure}

We use a separate transfer stage to maximize the transfer efficiency from the blue to the green MOT. This bridges the highly different operation regimes in temperature, magnetic field gradient and size due to the more than two orders of magnitude smaller linewidth of the green transition. 
For the transfer stage we apply the well known method of frequency-broadening the narrow-linewidth MOT light~\cite{Yb_first_MOT_Kuwamoto, Yb_first_dispenser_2DMOT_Dorscher, Sr_BEC_Schreck, Yb_core-shell_MOT_Mun} to increase both the capture velocity and the capture volume. We drive an acousto-optical modulator with a signal frequency-modulated by a sawtooth waveform of \qty{1}{\micro\second} period, shown in Fig.~\ref{fig: Green MOT transfer}(a). The modulation is controlled by an in-house developed Direct Digital Synthesizer-based programmable signal generator.
During the \qty{60}{\milli\second} transfer stage, we linearly decrease the amplitude of the frequency broadening to zero. Simultaneously, we ramp down the green detuning, the blue and green powers and the magnetic field gradient, as shown in Fig.~\ref{fig:2D-3D view of the setup}(c).

The relevant parameters for the optimization of the frequency modulation are defined in Fig.~\ref{fig: Green MOT transfer}(a): the start frequency $f_a$, the end frequency $f_b$, the sweep center frequency $f_c = (f_b+f_a)/2$ and the sweep range $\delta_f = f_b - f_a$. The sign of the range $\delta_f$ defines the sweep direction with $\delta_f >0$ for an upward sweep and $\delta_f <0$ for a downward sweep. The resonance frequency of the green transition $f_0$ is denoted by the black dashed line.
Figure~\ref{fig: Green MOT transfer}(b) shows the final atom number $N_\text{A}$ as a function of the sweep range $\delta_f$ and the detuning of the center frequency $f_c-f_0$. 
The black dashed line denotes the condition $f_c - f_0 = -|\delta_f /2|$, at which the atomic resonance is reached at the upper edge of the sweep range.
 We observe a significant improvement in the atom number $N_\text{A}$ for a wide spread of parameters, compared to the case at $\delta_f =0$ with no frequency broadening. Such enhancement predominantly occurs when the entire sweep remains in the negative-detuning region (left of the black dashed lines). 
 Furthermore, we find that the enhancement in this region is symmetric around $\delta_f =0$, indicating that the sweep direction does not play a role. 
 However, we observe a clear asymmetry in the region right of the black dashed lines, where the sweep crosses the atomic resonance $f_0$. We interpret this as the so called sawtooth-wave adiabatic passage (SWAP) cooling mechanism~\cite{Swap_cooling_Holland, Swap_MOT_Holland} observed for Strontium~\cite{norcia_narrow-line_2018, Sr_swap_cooling_Blatt, Sr_Steady_state_MOT_Schreck} and Dysprosium~\cite{Dy_SWAPCooling_Windpasinger}, where upward frequency sweeps across the resonance cause absorption and stimulated emission of photons from counter-propagating beams which increases the MOT forces.
 Nevertheless, in our system SWAP cooling is less efficient than simple negatively-detuned frequency broadening, probably because the green transition is still broader than the Strontium narrow transition for which SWAP cooling has been extensively optimized~\cite{Sr_swap_cooling_Blatt}.
 
For our transfer stage we choose the parameters $\delta_f \approx 50\times\Gamma_{556}$ and $f_c-f_0 \approx -50\times\Gamma_{556}$.
To estimate the transfer efficiency of the blue-green handover we use a transfer-and-recapture method. After loading the blue MOT, we transfer the atoms to the green MOT and immediately recapture into the blue MOT. From the comparison of the blue fluorescence before the transfer stage and after the recapture, we determine an efficiency of $80\%$.
With this transfer scheme we obtain a final atom number $N_\text{A}$ a factor of \qty{\sim2}{} higher than the best transfer without frequency broadening ($\delta_f =0$ row in Fig.~\ref{fig: Green MOT transfer}(b)), and a factor of \qty{\sim10}{} higher compared to no transfer stage at all, measured separately by reducing the transfer duration to zero.

\section{\label{sec:GreenMOT}Green 3D MOT}

\begin{figure}
    \centering
    \includegraphics[scale = 1]{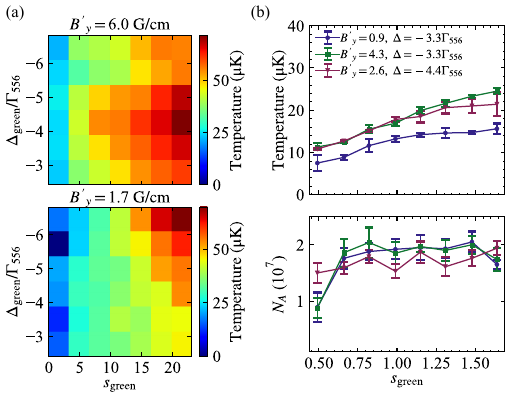} 
    \caption{\label{fig:3D Green MOT characterization}
        Green 3D MOT optimization. (a) Temperature of the atoms from time-of-flight imaging at $B'_y=\qty{6.0}{\gauss\per\centi\meter}$ (top) and \qty{1.7}{\gauss\per\centi\meter} (bottom), for a coarse scan of saturation parameter $s_\text{green}$ and detuning $\Delta_\text{green}$. (b) Finer scan of the low intensity range for selected detunings and magnetic field gradients, showing   temperature (top) and atom number (bottom) of the green MOT. The temperature of the atom cloud reaches below \qty{10}{\micro\kelvin} without atom loss.
    }
\end{figure}

After the transfer stage, we end our experiment sequence by holding the atoms in the green 3D MOT for \qty{100}{\milli\second}, during which they reach thermal equilibrium and a steady-state spatial distribution. We optimize the final green MOT parameters to achieve the lowest temperature without significant atom loss, namely magnetic field gradient $B'_y$, detuning $\Delta_\text{green}$ and saturation parameter $s_\text{green}$.
For two exemplary magnetic field gradients, the two panels of Figure~\ref{fig:3D Green MOT characterization}(a) show the temperature of the atom cloud measured by time-of-flight imaging. We observe that the temperature depends moderately on the detuning and strongly on the saturation parameter, while the atom number remains constant within the measurement uncertainty over both scans (not shown in the figure). At low saturation parameter and higher magnetic field gradient (top panel) the temperature reaches \qty{\sim25}{\micro\kelvin}, while, at low magnetic field gradient (lower panel) we reach \qty{\sim14}{\micro\kelvin}.

To further approach the Doppler temperature of \qty{4.4}{\micro\kelvin}, we explore the range of even lower saturation parameters ($s_\text{green}<1$) and magnetic field gradients in Figure~\ref{fig:3D Green MOT characterization}(b).
The downward trend in temperature continues while the atom number remains constant until  $s_\text{green} = 0.65$, below which atom loss sets in. 
With parameters $s_\text{green} = 0.65$, $\Delta_\text{green} = -3.3\times\Gamma_{556}$ and $B'_y = \qty{0.9}{\gauss\per\centi\meter}$, we achieve a temperature of $T_A \approx \qty{10}{\micro\kelvin}$ with an atom number of $N_\text{A} \approx 1.8\times10^7$.

\section{\label{sec7:level 1: conclusions}Conclusions}
    
In this work, we have presented a compact setup for trapping and cooling Ytterbium atoms. By using a permanent-magnet-based 2D MOT we avoid large Zeeman slowers and their water-cooled magnetic coils. Furthermore, our simple optical setup, with the broadband and the narrow-linewidth MOTs separated in time instead of in space, offers long term stability and less parameters to optimize, compared to more complex schemes such as the core-shell MOT or spatially separated slowing beams.
Additionally, we found that shelving into the long-lived ${}^3P_1$ state is beneficial for increasing the number of atoms trapped in Yb MOTs.
The optimization of the setup for $\mathrm{{}^{174}Yb}$ leads to trapping $\sim1.8\times10^7$ atoms at about twice the narrow-line Doppler temperature within \qty{2.5}{\second}.
The number of trapped atoms could be enhanced by applying a sideband broadening method in the 2D MOT demonstrated in Ref.~\cite{barbiero_sideband-enhanced_2020}.
Further improvements of green power stabilization far below saturation intensity and precise compensation of stray magnetic fields at low gradients may enable cooling even closer to the Doppler limit.
Our two-chamber apparatus with large optical access and stable ultra-high vacuum, together with the demonstrated experimental scheme, sets the basis for further optical trapping and evaporative cooling to quantum degeneracy as well as for nonlinear quantum optics experiments with Rydberg atoms~\cite{firstenberg_nonlinear_2016, murray_chapter_2016, pritchard_cooperative_2010, dudin_strongly_2012, peyronel_quantum_2012}.

\textit{Acknowledgements} We thank Phillip Lunt, Simon .W. Ball, Mohammad Noaman, and Rafael Rothganger de Paiva for contributions to the experimental apparatus and Onno Smit for contributions to the numerical simulations. This work was supported by the European Union's Horizon 2020 program under the ERC grant SUPERWAVE (grant No.101071882) and by the Deutsche Forschungsgemeinschaft (DFG) within the collaborative research center SFB/TR185 OSCAR, project A8 (No. 277625399). 

\textit{Data availability} The experimental data presented in this work is available in the repository \href{https://doi.org/10.5281/zenodo.13177244}{10.5281/zenodo.13177244}.


\begin{appendices}

\section{Numerical simulations of the 2D-3D MOT system}
\label{sec:MOT-simulation}

This appendix describes simulations of atom dynamics in the 2D-3D MOT system presented in the main text. 
Specifically, we calculate trajectories of atoms from the Yb dispensers through the 2D MOT into the (blue) 3D MOT.
This allows us to calculate the expected dependence of the fraction of captured atoms on parameters which are optimized in the experiment.

We use PyLCP, a Python package for laser cooling~\cite{ECKEL2022108166}, to implement the radiation force and numerically integrate the equations of motion.
The force at each position is evaluated based on the local magnitudes and orientations of the magnetic and optical vector fields.
We reproduce the geometry and the experimental parameters of the optimization of the 2D MOT presented in the main text (cf.~Sec.~\ref{sec4:level 1: Yb atom source}).
We then compare the forces calculated using different models implemented in the package, namely the rate equation, the optical Bloch equation, and the heuristic model. We find that the heuristic model shows a good compromise between physical accuracy and computational efficiency for the simulation of the blue atomic transition at \SI{\sim399}{\nano\meter}.
For our 2D and 3D MOT beams, we modify the heuristic model (Eq.~(35) in Ref.~\cite{ECKEL2022108166}) such that the radiation pressure force calculated for each beam considers the saturation by this beam alone instead of the summed-up saturation by all beams.
Specifically, we replace the summed saturation term in the denominator by the saturation parameter of only the single beam.
This modification greatly improves the agreement of the heuristic model with the full Bloch equation model, and it is physically justified by the fact that for most of the MOT region only one of the beams is near-resonant and significantly excites the atoms, while the other beams contribute almost nothing to the saturation due to the large magnetic fields and high detuning used at our typical parameters.
For the low intensity push beam, however, we consider the MOT beams for saturation, i.e.\ the saturation term in the denominator consists of the sum of the saturation parameters of all beams. 
We benchmark this modified model by comparing the radiation force profiles with the results given by the rate equation model.

\begin{figure}
   \centering
   \includegraphics[scale = 1]{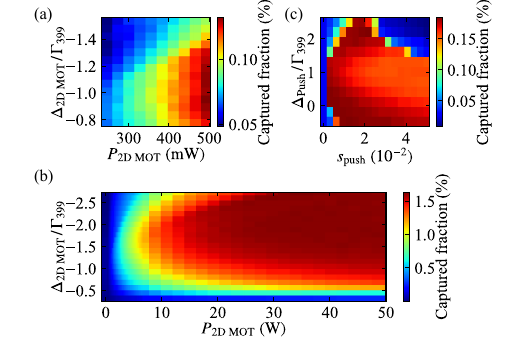}
    \caption{
    \label{fig:sim_plots}
    Simulation results of the 2D MOT and push beam parameter scan. We use all experimental parameters as for the data presented in Fig.~\ref{fig:2D MOT performance}(a) and~(b). Shown is the fraction of captured atoms from the total number of sampled ones. (a)~2D MOT parameter scan with the parameter range as in Fig.~\ref{fig:2D MOT performance}(a). (b) 2D MOT parameter scan over a larger range unavailable in the experiment. (c) Push beam parameter scan.
    }
\end{figure}

Initial conditions of $\sim10^5$ atoms are randomly sampled from a Maxwell-Boltzmann distribution with a temperature of \SI{500}{\kelvin}, starting from a point located on one of the four dispensers. 
The angles are constrained by the dispenser shielding plates (see Fig.~\ref{fig:2DMOT Cross Section}).
Whether a particle is captured by the blue 3D MOT is determined by whether it passes the differential pumping tube and whether its velocity gets close to zero at the center of the MOT.
Figure~\ref{fig:sim_plots} shows the simulated scan of the 2D MOT total power and detuning, as well as the scan of the push beam intensity and detuning.
The simulation of the 2D MOT parameters shown in Fig.~\ref{fig:sim_plots}(a), with \SI{0.12}{\percent} maximum captured fraction, reproduces the trend of the optimization measurements presented in Fig.~\ref{fig:2D MOT performance}(a). 
As the flux from our 2D MOT is limited by the available optical power in the experiment, with the simulation we explore a higher range up to \SI{50}{\watt} in Fig.~\ref{fig:sim_plots}(b). The number of captured atoms saturates at a few tens of Watts of total power, trapping up to \SI{1.6}{\percent} of the hot atoms emitted by the dispenser.

The scan of the push beam parameters in Fig.~\ref{fig:sim_plots}(c) shows that we can find parameter values which significantly enhance the number of captured atoms, compared to the case without the push beam corresponding to $s_\mathrm{push}=0$. We observe similar improvements for both positive-detuned and negative-detuned push beam frequencies.
The region with low capture fraction at large positive detunings and powers, on the top right of Fig.~\ref{fig:sim_plots}(c), is due to atoms being accelerated to velocities higher than the capture velocity of the 3D MOT.
However, this simulation result for the push beam differs from our experimental observation of a single optimum on the positively detuned side. 
In fact, we find that simulation results of the push beam parameter scan are very sensitive to the exact geometry and initial parameters.
Real experimental conditions, e.g.\ small deviations of the 2D MOT axis from a straight line (cf.\ Sec.~\ref{sec4:level 1: Yb atom source}), cannot be easily reproduced in the simulations, but they may act as effective velocity filters restraining the effective push beam parameter range.

The simulation can be extended to also model the green shelving during the blue MOT, the transfer stage as well as the green 3D MOT by making the magnetic and optical fields time-dependent.
However, simulation including the narrow green transition at \SI{556}{\nano\meter} requires consideration of recoil, and the heuristic model may not be accurate enough, making such simulations computationally intensive.

\end{appendices}
\newpage
\bibliography{YBMOT}

\end{document}